\documentclass[12pt]{revtex4}
\usepackage{url}
\usepackage{graphicx}
\begin{document}

\title{Role of diversification risk in financial bubbles}

\author{Wanfeng Yan $\dag$, Ryan Woodard $\dag$ and Didier Sornette $\dag \ddag$}
\email{wyan@ethz.ch, rwoodard@ethz.ch, dsornette@ethz.ch}
\thanks{\\~Corresponding author: Didier Sornette.}
\address{\normalfont{$\dag$ Chair of Entrepreneurial Risks,\\
Department of Management, Technology and Economics,\\ ETH Z\"{u}rich,
CH-8001 Z\"{u}rich, Switzerland. \vspace{12pt}\\
$\ddag$ Swiss Finance Institute,\\ c/o University of Geneva, 40 blvd. Du Pont
dArve, CH-1211 Geneva 4, Switzerland.}\\\vspace{12pt}}

\newpage

\begin{abstract}
We present an extension of the Johansen-Ledoit-Sornette (JLS) model to include
an additional pricing factor called the ``Zipf factor'', which describes the
diversification risk of the stock market portfolio. Keeping all the dynamical
characteristics of a bubble described in the JLS model, the new model provides
additional information about the concentration of stock gains over time. This
allows us to understand better the risk diversification and to explain the
investors' behavior during the bubble generation. We apply this new model to
two famous Chinese stock bubbles, from August 2006 to October 2007 (bubble 1)
and from October 2008 to August 2009 (bubble 2). The Zipf factor is found
highly significant for bubble 1, corresponding to the fact that valuation gains
were more concentrated on the large firms of the Shanghai index. It is likely
that the widespread acknowledgement of the 80-20 rule in the Chinese media and
discussion forums led many investors to discount the risk of a lack of
diversification, therefore enhancing the role of the Zipf factor. For bubble 2,
the Zipf factor is found marginally relevant, suggesting a larger weight of
market gains on small firms. We interpret this result as the consequence of the
response of the Chinese economy to the very large stimulus provided by the
Chinese government in the aftermath of the 2008 financial crisis.
\end{abstract}

\maketitle

\clearpage

\section{Introduction}

Johansen et al. \cite{js,jsl,jls} developed a model (referred to below as the JLS
model) of financial bubbles and crashes, which is an extension of
the rational expectation bubble model of Blanchard and Watson \cite{Blanchardwat}. In
this model, a crash is seen as an event potentially terminating the
run-up of a bubble. A financial bubble is modeled as a regime of
accelerating (super-exponential power law) growth punctuated by
short-lived corrections organized according to the symmetry of discrete
scale invariance \cite{DSI-sornette98}. The super-exponential power
law is argued to result from positive feedback resulting from noise
trader decisions that tend to enhance deviations from fundamental
valuation in an accelerating spiral.

The JLS model has been proved to be a very powerful and flexible tool to detect
financial bubbles and crashes in various kinds of markets such as the 2006 -
2008 oil bubble \cite{oil}, the Chinese index bubble in 2009
\cite{Jiangetal09}, the real estate market in Las Vegas \cite{vegas}, the South African
stock market bubble \cite{southafrica} and the US repurchase agreement market
\cite{repo}. Recently, the JLS model has been extended to detect market
rebounds \cite{rebound} and to infer the fundamental market value hidden
within observed prices \cite{pp1}. Also, new
experiments in ex-ante bubble detection and forecast has been performed in
the Financial Crisis Observatory at ETH Zurich \cite{BFEFCO09, BFEFCO10}.

Here,  we present an extension of the JLS model, which is in the spirit of the
approach developed by Zhou and Sornette \cite{zhouSornette06} to include
additional pricing factors.

The literature on factor models is huge and we refer e.g. to
Ref.\cite{KnightSatchell} and references therein for a review of the
literature. One of the most famous factor model, now considered as a standard
benchmark, is the three-factor Fama-French model
\cite{FamaandFrench1992,FamaandFrench1993,FamaandFrench1995,FamaandFrench1996}
augmented by the momentum factor \cite{Carhart1997}.
Recently, the concept of the Zipf factor has been introduced
\cite{MalevergneSortwofactor,MalevergneSantaSor}. The key idea of the Zipf
factor is that, due to the concentration of the market portfolio when the
distribution of the capitalization of firms is sufficiently heavy-tailed as is
the case empirically, a risk factor generically appears in addition to the
simple market factor, even for very large economies. Malevergne et al.
\cite{MalevergneSortwofactor,MalevergneSantaSor} proposed a simple proxy for
the Zipf factor as the difference in returns between the equal-weighted and the
value-weighted market portfolios.  Malevergne et al.
\cite{MalevergneSortwofactor,MalevergneSantaSor} have shown that the resulting
two-factor model (market portfolio $+$ the new factor termed ``Zipf factor'')
is as successful empirically as the three-factor Fama-French model.
Specifically, tests of the Zipf model with size and book-to-market
double-sorted portfolios as well as industry portfolios finds that the Zipf
model performs practically as well as the Fama-French model in terms of the magnitude and
significance of pricing errors and explanatory power, despite that it has only
two factors instead of three.

In the present paper, we would like to introduce a new model by combining the
Zipf factor with the JLS model. The new model keeps all the dynamical
characteristics of a bubble described in the JLS model. In addition, the new
model can also provide the information about the concentration of stock gains
over time from the knowledge of the Zipf factor. This new information is very
helpful to understand the risk diversification and to explain the investors'
behavior during the bubble generation.

The paper is constructed as follows. Section \ref{sec:model} describes the
definition of the Zipf factor as well as the new model. The derivation of the
model is presented in this section and the appendix. Section
\ref{sec:calibration} introduces the calibration method of this new model. Then
we test the new model with two famous Chinese stock bubbles in the history in
Section \ref{sec:application} and discuss the role of the Zipf factor in these
two bubbles. Section \ref{sec:conclusion} concludes.

\section{The model}
\label{sec:model}

We introduce the new model in this section. Our goal is to combine the Zipf
factor $z(t) dt$ with the JLS model of the bubble dynamics. To be specific, we
introduce the following definition.\\

{\bf Definition 1}: {\it The Zipf factor $z(t) dt$ is defined as proportional
to the difference between the returns of the capitalization-weighted portfolio
and the equal-weighted portfolio for the last time step:
\begin{equation}
  z(t) dt :=  \frac{dp }{p(t)} -   \frac{dp_e}{p_e(t)}~,
  \label{hjyjuj4u}
\end{equation}
where   $p$ (respectively $p_{e}$) is
the price of the capitalization-weighted (respectively
equal-weighted) portfolio, $dp := p(t) - p(t-dt)$ and $dp_e := p_e(t) - p_e(t-dt)$.
The weights of the portfolios are normalized so that their two prices are identical at the day
preceding the beginning time $t_0$ of the time series: $p_e(t_0) =
p(t_0)$.} \\

{\bf Definition 2}: {\it The  integrated Zipf factor $\zeta(t)$ is obtained by taking
the integral of the Zipf factor defined by expression (\ref{hjyjuj4u}):
\begin{equation}
\zeta(t) := \ln p(t) - \ln p_e(t)~.
\label{heyjujkuj5}
\end{equation}
\it}

\noindent By definition, the Zipf factor describes the exposition to
a lack of diversification due to the concentration of the stock
market on a few very large firms.

The dynamics of stock markets
during a bubble regime is then described as
\begin{equation}
  \frac{dp(t)}{p(t)} = \mu(t)dt + \gamma z(t)dt + \sigma(t)dW - \kappa dj~,
  \label{eq:dynamic}
\end{equation}
where $p$ is the portfolio price, $\mu$ is the drift (or trend) whose
accelerated growth describes the presence of a bubble (see below),
$\gamma$ is the factor loading on the Zipf's factor and
$dW$ is the increment of a Wiener process (with zero mean and unit
variance). The term $dj$
represents a discontinuous jump such that $dj = 0$ before the crash
and $dj = 1$ after the crash occurs. The loss amplitude associated
with the occurrence of a crash is determined by the parameter
$\kappa$. The assumption of a constant jump size is easily relaxed
by considering a distribution of jump sizes, with the condition that
its first moment exists. Then, the no-arbitrage condition is
expressed similarly with $\kappa$ replaced by its mean. Each
successive crash corresponds to a jump of $dj$ by one unit. The
dynamics of the jumps is governed by a crash hazard rate $h(t)$.
Since $h(t) dt$ is the probability that the crash occurs between $t$
and $t+dt$ conditional on the fact that it has not yet happened, we
have $E_t[dj]  = 1 \times h(t) dt + 0 \times (1- h(t) dt)$, where $E_t[.]$ denotes
the expectation operator. This leads to
\begin{equation}
  {\rm E}_t[dj] = h(t)dt~.
  \label{theyjytuj}
\end{equation}

Noise traders exhibit collective herding behaviors that may
destabilize the market in this model. We assume that the aggregate
effect of noise traders can be accounted for by the following
dynamics of the crash hazard rate
\begin{equation}
  h(t) = B'(t_c-t)^{m-1}+C'(t_c-t)^{m-1}\cos(\omega\ln (t_c-t) -\phi')~.
  \label{eq:hazard}
\end{equation}
The intuition behind this specification (\ref{eq:hazard}) has been
presented at length by Johansen et al. \cite{js,jsl,jls}, and
further developed by Sornette and Johansen \cite{SorJohansenQF01},
Ide and Sornette \cite{IdeSornette} and Zhou and Sornette
\cite{zhouSornette06}. In a nutshell, the power law behavior $\sim
(t_c-t)^{m-1}$ embodies the mechanism of positive feedback posited
to be at the source of the bubbles. If the exponent $m<1$, the crash
hazard may diverge as $t$ approaches a critical time $t_c$,
corresponding to the end of the bubble. The cosine term in the
r.h.s. of (\ref{eq:hazard}) takes into account the existence of a
possible hierarchical cascade of  panic acceleration punctuating the
course of the bubble, resulting either from a preexisting hierarchy
in noise trader sizes \cite{SornetteJohansen97} and/or from the
interplay between market price impact inertia and nonlinear
fundamental value investing  \cite{IdeSornette}.

We assume that all the investors of the market have already taken
the diversification risk into account, so that the no-arbitrage
condition reads ${\rm E}_t[\frac{dp(t)}{p(t)}- \gamma z(t) dt]=0$,
where the expectation is performed with respect to the risk-neutral
measure, and in the frame of the risk-free rate. This is the
condition that the price process concerning the diversification risk
should be a martingale. Taking the expectation of expression
(\ref{eq:dynamic}) under the filtration (or history) until time $t$
reads
\begin{equation}
{\rm E}_t \left[\frac{dp}{p} - \gamma z dt \right]= \mu(t) dt + \sigma(t)  {\rm E}_t[dW] - \kappa {\rm E}_t[dj]~.
 \label{thetyjye}
\end{equation}
Since ${\rm E}_t[dW] =0$ and  ${\rm E}_t[dj] = h(t)dt$ (equation
(\ref{theyjytuj})), together with the no-arbitrage condition ${\rm
E}_t[dp(t)]=0, \forall t$, this yields
\begin{equation}
\mu(t) = \kappa h(t)~.
\label{tjyj4n}
\end{equation}
This result (\ref{tjyj4n}) expresses that the return $\mu(t)$ is controlled by the risk of the crash
quantified by its crash hazard rate $h(t)$. The
excess return $\mu(t) = \kappa h(t)$ is the remuneration that
investors require to remain invested in the bubbly asset,  which is
exposed to a crash risk.

Now, conditioned on the fact that no crash occurs,
equation (\ref{eq:dynamic}) is simply
\begin{equation}
  \frac{dp(t)}{p(t)} - \gamma z(t) = \mu(t)dt + \sigma(t)dW = \kappa h(t) dt + \sigma(t)dW~,
  \label{eq:dynaiuhomic}
\end{equation}
where the Zipf factor $z(t)$ is given by expression (\ref{hjyjuj4u}).
Its conditional expectation leads to
\begin{equation}
{\rm E}_t \left[\frac{dp(t)}{p(t)} - \gamma z(t) \right] = \kappa h(t) dt
\end{equation}
Substituting with the expression (\ref{eq:hazard}) for $h(t)$ and  (\ref{hjyjuj4u}) for $z(t)$,
and integrating, yields the log-periodic power law (LPPL) formula as in the JLS model, but
here augmented by the presence of the Zipf factor, which adds the term proportional to the
Zipf factor loading $\gamma$:
\begin{equation}
  {\rm E}_t[\ln p(t) - \gamma \zeta(t)] = A + B(t_c-t)^m + C(t_c-t)^m\cos(\omega\ln (t_c-t) - \phi)~,
\label{eq:lppl}
\end{equation}
where $\zeta(t)$ is defined by expression (\ref{heyjujkuj5}) and the r.h.s. of (\ref{eq:lppl})
is the primitive of expression (\ref{eq:hazard}) so that
$B = - \kappa B' /m$ and $C = - \kappa C' /
\sqrt{m^2+\omega^2}$. This expression (\ref{eq:lppl})
describes the average price dynamics only up to the end of the
bubble. The same structure as equation (\ref{eq:lppl}) is
obtained using a stochastic discount factor following
the derivation of Zhou and Sornette \cite{zhouSornette06}, as shown in the appendix.

The JLS model does not specify what happens beyond $t_c$.
This critical $t_c$ is the termination of the bubble regime and the
transition time to another regime. This regime could be a big crash
or a change of the growth rate of the market. Merrill Lynch EMU
(European Monetary Union) Corporates Non-Financial Index in 2009
\cite{BFE-FCO09} provides a vivid example of a change of regime
characterized by a change of growth rate rather than by a crash or rebound.
 For $m<1$, the crash hazard rate accelerates up to $t_c$ but
its integral up to $t$ which controls the total probability for a
crash to occur up to $t$ remains finite and less than $1$ for all
times $t \leq t_c$. It is this property that makes it rational for
investors to remain invested knowing that a bubble is developing and
that a crash is looming. Indeed, there is still a finite probability
that no crash will occur during the lifetime of the bubble. The condition that the price remains finite
at all time, including $t_c$, imposes that $m > 0$.

Within the JLS framework, a bubble is qualified when
the crash hazard rate accelerates. According to (\ref{eq:hazard}), this
imposes $m<1$ and $B'>0$, hence $B<0$ since $m > 0$ by the condition
that the price remains finite. We thus have a first condition for a bubble to occur
\begin{equation}
  0 < m < 1~.
\label{eq:m}
\end{equation}
By definition, the crash rate should be non-negative. This imposes \cite{bm}
\begin{equation}
 b \equiv -Bm - |C|\sqrt{m^2+\omega^2}  \geq 0~.
  \label{eq:bg0}
\end{equation}

\section{Calibration method}
\label{sec:calibration}

There are eight parameters in this LPPL model augmented by the introduction of
the Zipf's factor,  four of which are the linear parameters ($\gamma, A, B$ and
$C$). The other four ($t_c, m, \omega$ and $\phi$) are nonlinear parameters.

We first slave the linear parameters to the nonlinear ones. The
method here is the same as used by Johansen et al.  \cite{jls}. The detailed
equations and procedure is as follows.
We rewrite Eq. (\ref{eq:lppl}) as:
\begin{equation}
  {\rm E}[\ln p(t)] = \gamma \zeta(t) + A + B f(t) + C g(t) := RHS(t)~.
 \label{eq:lppl2sgw}
\end{equation}
We have also defined
\begin{equation}
f(t) = (t_c-t)^m~,~~~~~  g(t) = (t_c-t)^m\cos(\omega\ln (t_c-t) - \phi)~.
\end{equation}

The minimization of the sum of the squared residuals should satisfy
\begin{equation}
  \frac{\partial \Sigma_t [\ln p(t) - RHS(t)]^2}{\partial \theta} = 0,~~~ \forall~ \theta \in \{\gamma, A, B, C\}.
\end{equation}
The linear
parameters $\gamma, A, B$ and $C$ are determined as the solutions of
the linear system of four equations:
\begin{equation}
  \Sigma_{t=t_1}^{t_2} \left(\begin{array}{cccc}
\zeta^2(t)&\zeta(t)&\zeta(t)f(t)&\zeta(t)g(t) \\ \zeta(t)&1&f(t)&g(t) \\ \zeta(t)f(t)&f(t)&f^2(t)&f(t)g(t) \\
\zeta(t)g(t)&g(t)&f(t)g(t)&g^2(t)
\end{array} \right) \left(\begin{array}{c} \gamma
\\ A \\ B \\ C \end{array} \right) = \Sigma_{t=t_1}^{t_2} \left(
\begin{array}{c}
\zeta(t) \ln p(t)  \\
\ln p(t) \\
f(t) \ln p(t) \\
g(t) \ln p(t) \end{array} \right) ~.
\end{equation}
This provides four analytical expressions for the four linear
parameters $(\gamma, A, B, C)$ as a function of the remaining
nonlinear parameters $t_c, m, \omega, \phi$. The resulting cost
function (sum of square residuals) becomes function of just the four
nonlinear parameters $t_c, m, \omega, \phi$. This achieves a very
substantial gain in stability and efficiency as the search space is
reduced to the 4 dimensional parameter space $(t_c, m, \omega,
\phi)$. A heuristic search implementing the taboo algorithm
\cite{ck} is used to find initial estimates of the parameters which
are then passed to a Levenberg-Marquardt algorithm \cite{kl,dm} to
minimize the residuals (the sum of the squares of the differences)
between the model and the data.  The calibration is performed for
the time window delineated by $[t_1, t_2]$, where $t_1$ is the
starting time and $t_2$ is the ending time of the price time being
fitted by expression (\ref{eq:lppl}) or equivalently
(\ref{eq:lppl2sgw}).

The bounds of the search space are:
\begin{eqnarray}
  t_c &\in& [t_2, t_2 + 0.375(t_2 - t_1)]\\\label{eq:rangetc}
  m &\in& [10^{-5}, 1-10^{-5}] \label{eq:rangem}\\
  \omega &\in& [0.01, 40] \\
  \phi &\in& [0, 2\pi - 10^{-5}]
\end{eqnarray}
We choose these bounds because $m$ has to be between $0$ and $1$
according to the discussion before; the log-angular frequency
$\omega$ should be greater than $0$. The upper bound $40$ is large
enough to catch high-frequency oscillations (though we later discard
fits with $\omega > 20$); the phase $\phi$ should be between 0 and
$2\pi$; The predicted critical time $t_c$ should be after the end
$t_2$ of the fitted time series. Finally, the upper bound of the
critical time $t_c$ should not be too far away from the end of the
time series since predictive capacity degrades far beyond $t_2$.
Jiang et al.  \cite{Jiangetal09} have found empirically that a
reasonable choice is to take the maximum horizon of predictability
to extent to about one-third of the size of the fitted time window.

\section{Application to the Shanghai Composite Index (SSEC)}
\label{sec:application}

\subsection{Construction of the capitalization-weighted and equally-weighted portfolios}

We use the Shanghai Composite Index as the market proxy to test the
JLS model augmented with the Zipf factor. The Shanghai Composite
Index is a capital-weighted measure of stock market performance. On
December 19, 1990, the base value of the Shanghai Composite Index
$I$ was fixed to $100$. We note the base date as $t_B$. Denoting by
$K_B$, the total market capitalization of the firms entering in the
Shanghai Composite index on $t_B$ December 19, 1990, the value
$p(t)$ of the Shanghai Composite Index at any later time $t$ is
given by
\begin{equation}
  p(t) = \frac{K(t)}{K_B} \times 100,
\end{equation}
where $K(t)$ is the current total market capitalization of the
constituents of the Shanghai Composite index. Here, time is counted
in units of trading days. Calling $p_j(t)$ (respectively $s_j(t)$),
the share price (respectively total number of shares) of firm $j$ at
time $t$, we have the total capitalization of firm $j$ at time $t$
\begin{equation}
  K_j(t) = p_j(t) s_j(t)~,
\end{equation}
and the total market capitalization at time $t$
\begin{equation}
  K(t) = \sum_{j=1}^{M(t)} K_j(t)~,
\end{equation}
where $M(t)$ is the number of the stocks listed in the index at time
$t$.

At the time when the calibrations were performed, the SSEC market
included 884 active stocks. Since December 19, 1990, 36 firms were
delisted and another 11 were temporarily stopped. Based on the rule
of the index calculation, the terminated stocks are deleted from the
total market capitalization after the termination is executed, while
the last active capitalization of the temporarily stopped stocks are
still included in the total market capitalization.

The equal-weighted price $p_e$ entering in the definition of the
Zipf factor is constructed according to the formula:
\begin{equation}
p_e(t) = p(t_0) \times \exp \left[\sum_{i=t_1}^t  r_e(i)\right]~,
\label{rtjuki8klo8k}
\end{equation}
where $t_1$ is the beginning of the fitted window and $t_0$ is the
trading day immediately preceding $t_1$. We use this measure of $p_e$ to make
sure that the equal-weighted price and the value-weighted price are
identical at $t_0$. This implies that $\zeta(t_0)$ is set to be $0$ (recall
that $\zeta$ is defined by expression (\ref{heyjujkuj5})).
The return $r_e(i)$ is defined by
\begin{equation}
r_e(i) = {1 \over M(i)}  \sum_{j=1}^{M(i)}  \left[ \ln K_j(i) - \ln
K_j(i-1) \right]~. \label{rheyju6h}
\end{equation}
In expression  (\ref{rheyju6h}), $K_j(i)$ is the total
capitalization value of firm $j$ at time $i$ and $M(i)$ is the
number of the stocks which are listed in the index for both time $i$
and $i-1$. Formula (\ref{rheyju6h}) together with
(\ref{rtjuki8klo8k}) means that the Zipf factor is a portfolio that
puts an equal amount of wealth at each time step (by a corresponding
dynamical reallocation depending on the relative performance of the
$M(i)$ stocks as a function of time) on each of the $M(i)$ stocks
entering in the definition of the Shanghai Composite Index, so that
the Zipf portfolio is maximally diversified (neglecting here the
impact of cross-correlations between the assets). Putting expression
(\ref{rheyju6h}) inside (\ref{rtjuki8klo8k}) yields
\begin{equation}
p_e(t) = p(t_0) \times \prod_{i=t_1}^t \left[
\left(\prod_{j=1}^{M(i)} { K_j(i) \over
K_j(i-1)}\right)^{1/M(i)}\right]~. \label{rtjuki8argwreklo8k}
\end{equation}
When the number of the stocks remains unchanged from $t_0$ to $t$,
i.e.
\begin{equation}
  M(i) = M,~~~~~ \forall i \in [t_0, t]~,
\end{equation}
expression (\ref{rtjuki8argwreklo8k}) can be simplified as:
\begin{equation}
p_e(t) = p(t_0) \times \left[\prod_{j=1}^M   \left({ K_j(t) \over
K_j(t_0)}\right)\right]^{1/M}~, \label{rtjuki8argwreklo8ksimp}
\end{equation}
showing that $p_e(t)$ is the geometrical mean of the capitalizations of the stocks
constituting the Shanghai Composite Index, as compared with the index
which is proportional to the arithmetic mean of the firm capitalizations.

\subsection{Empirical test of the JLS model augmented by the Zipf factor}

\begin{figure}
\centering
\includegraphics[width=\textwidth]{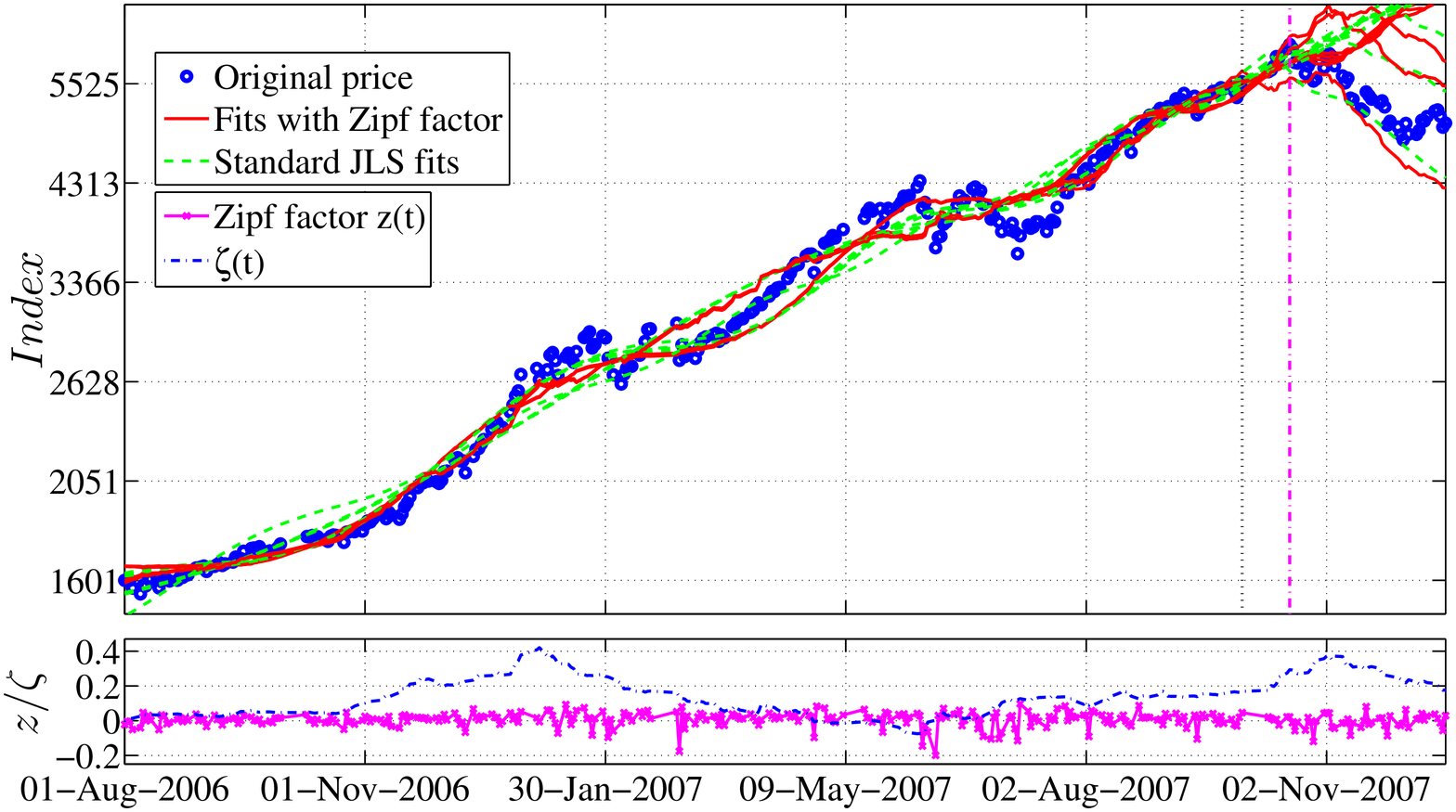}
\caption{Calibration of the new factor model and the original JLS
model to the Shanghai Composite Index (SSEC) between Aug-01-2006 and
Sep-28-2007. (Upper panel) The beginning of the fit interval is the
left boundary of the plot, while the end of the fit interval is
indicated by the vertical thick black dotted line. The real critical
time $t_c$ when the crash started is marked by the vertical magenta
dot-dashed line. The historical close prices are shown as blue full
circles. The best 10 fits of the original JLS model are shown as the
green dashed lines and the best 10 fits of the new factor model are
shown as the red solid lines. (Lower panel) The corresponding Zipf
factor (magenta solid line with `x' symbol) and $\zeta$ function
(blue dot-dashed line) during this period.}
\label{fg:fit_ssec_zipf1}
\end{figure}

\begin{figure}
\centering
\includegraphics[width=\textwidth]{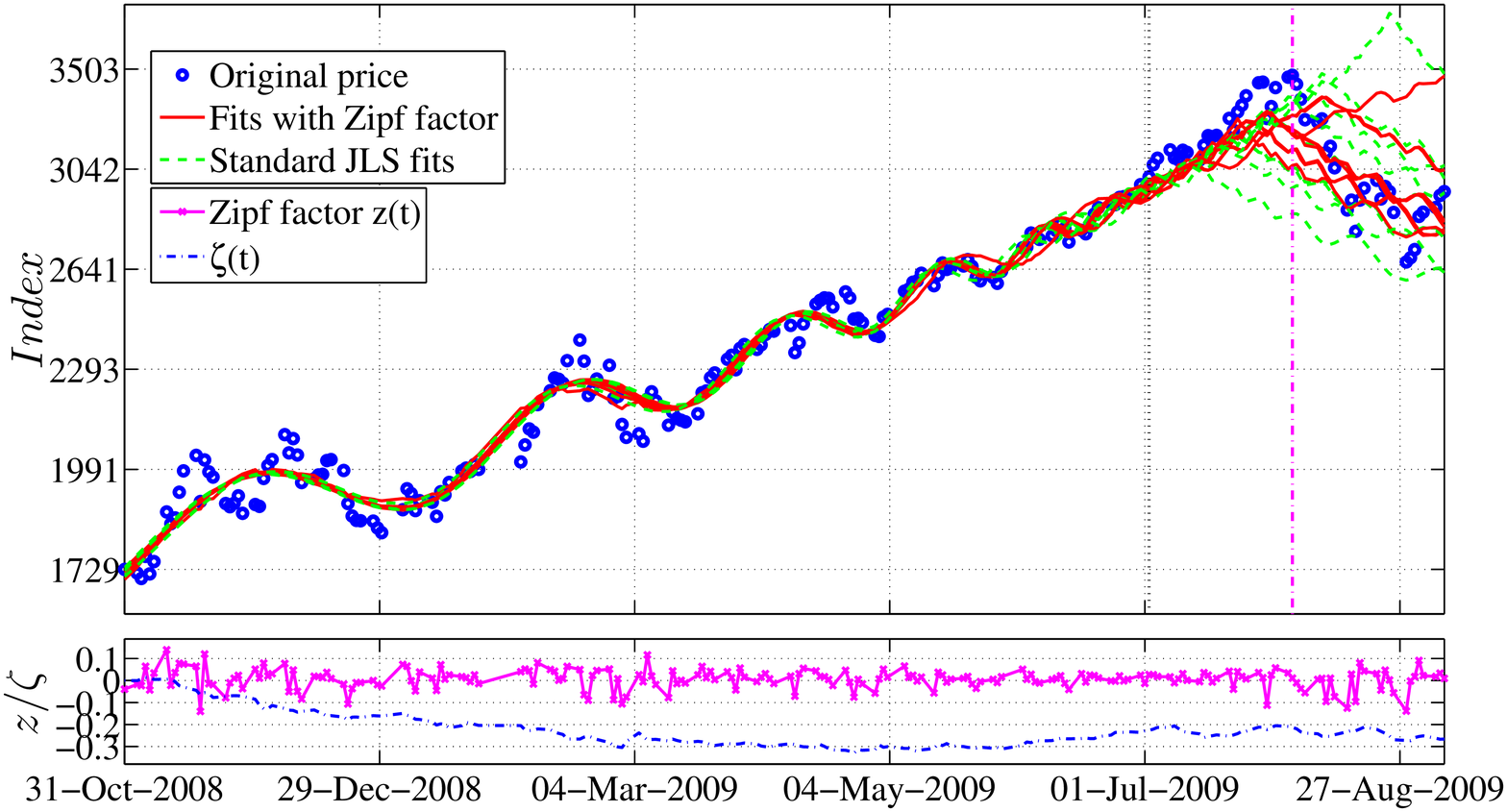}
\caption{Calibration of the new factor model and the original JLS
model to the Shanghai Composite Index (SSEC) between Oct-31-2008 and
Jul-01-2009. (Upper panel) The beginning of the fit interval is the
left boundary of the plot, while the end of the fit interval is
indicated by the vertical thick black dotted line. The real critical
time $t_c$ when the crash started is marked by the vertical magenta
dot-dashed line. The historical close prices are shown as blue full
circles. The best 10 fits of the original JLS model are shown as the
green dashed lines and the best 10 fits of the new factor model are
shown as the red solid lines. (Lower panel) The corresponding Zipf
factor (magenta solid line with `x' symbol) and $\zeta$ function
(blue dot-dashed line) during this period.}
\label{fg:fit_ssec_zipf2}
\end{figure}

The Shanghai Composite Index had two famous bubbles in recent history as
described in Table \ref{tb:ssecbubbles}. Both of them are tested in
this paper. The time series are fitted with both the original JLS
model and the new model. The 10 best initial guesses from the
heuristic search algorithm are kept. The results are shown in Figs.
\ref{fg:fit_ssec_zipf1} - \ref{fg:fit_ssec_zipf2}.

\begin{table}[h]
  \centering
    \begin{tabular}{|l|l|l|l|}
    \hline
    Example &Calibration start at $t_1$ & Prediction start at $t_2$ & Peak date of the bubble\\
    \hline
    Bubble 1 & Aug-01-2006 & Sep-28-2007 & 16-Oct-2007 \\
    Bubble 2 & Oct-31-2008 & Jul-01-2009 & Aug-04-2009 \\
    \hline
    \end{tabular}%
  \caption{Information on the tested bubbles of SSEC.}
  \label{tb:ssecbubbles}%
\end{table}%

We use the standard Wilks test of nested hypotheses to check the
improvement of the new factor model. This test assumes independent
and normally distributed residuals. The null hypothesis is:
\begin{enumerate}
\item[$H_0$: ] the original JLS model is sufficient and the new factor model is not necessary.
\end{enumerate}
The alternative hypothesis reads:
\begin{enumerate}
\item[$H_1$: ] The original JLS model is not sufficient and the new factor model is needed.
\end{enumerate}

For sufficiently large time windows, and noting $T$ the number of
trading days in the fitted time window $[t_1, t_2]$, the Wilks
log-likelihood ratio reads
\begin{equation}
W = 2 \log \frac{L_{Zipf,max}}{L_{JLS,max}} = 2T \ln
\frac{\sigma_{JLS}}{\sigma_{Zipf}} + \frac{\sum_{t=1}^T
R_{JLS}^2(t)}{\sigma_{JLS}^2} - \frac{\sum_{t=1}^T
R_{Zipf}^2(t)}{\sigma_{Zipf}^2}~, \label{ththwfq}
\end{equation}
where $R_{JLS}$ and $\sigma_{JLS}$ (respectively $R_{Zipf}$ and
$\sigma_{Zipf}$) are the residuals and their corresponding standard
deviation for the original JLS model (respectively the new factor
model).

In the large $T$ limit, and under the above conditions of asymptotic
independence and normality, the $W$-statistics is distributed with a
$\chi_k^2$ distribution with $k$ degrees of freedom, where $k$ is
the difference between the number of parameters in two models. In
our case, the new factor model has one more parameter, which is
$\gamma$. Therefore, $W$ in Eq.(\ref{ththwfq}) should follow the
$\chi_1^2$ distribution.

Only considering the best fit for each of the two models, we obtain a
$p$-value associated with the empirical value of the $W$-statistics
equal to $2.64 \times 10^{-7}$ for bubble 1 and $0.2517$ for bubble
2. Thus, the null hypothesis is rejected and the Zipf factor is
necessary for the best fit of bubble 1, while the null hypothesis is
not rejected and the Zipf factor is not necessary for the best fit
of bubble 2. This result is also consistent with the two values found for
$\gamma$, where $\gamma = 0.44$ for bubble 1 and $\gamma =
-0.028$ for bubble 2, showing the Zipf factor in bubble 1 plays an
important role in the improvement of the fit quality.

Keeping the best 10 fits as we described before increases the statistical power
of the Wilks test (simply by having more statistical data) and we want to show
that the new JLS model with the Zipf factor is an significant improvement. For
this, we combine all of the residuals from the best 10 fits to the data into a
large residual sample and calculate the Wilks log-likelihood ratio $W$ for this
large sample as defined by expression (\ref{ththwfq}). The corresponding
p-values are $0$ for bubble 1 and $0.0119$ for bubble 2. This means the new
factor model performs better than the original JLS model for both cases when we
consider the overall qualify of the best 10 fits.

A natural and interesting test is to find out if the new model with Zipf factor
has a better predictability of the critical time. To achieve this goal, two
examples are fitted by both models within different time windows obtained by
varying their start time $t_1$ and the end time $t_2$. We consider 15 different
values of $t_1$ and of $t_2$ in steps of 3 days, yielding 225 time series for
each example. We keep the best 10 fits for each time series and get 2250
predicted critical time $t_c$ with each model and for each example. The results
in Table \ref{tb:tcperform} show that the mean value and the standard deviation
of the critical time $t_c$ for both models are similar. The new model including
the Zipf factor neither improves nor deteriorates the predictability of the
critical time for these two examples.

\begin{table}[h]
  \centering
    \begin{tabular}{|l|l|l|l|}
    \hline
    Example & Peak date & Mean(std) of $t_c$, new model & Mean(std) of $t_c$, original model \\
    \hline
    Bubble 1 &16-Oct-2007&07-Oct-2007(55.6)&18-Oct-2007(54.1) \\
    Bubble 2 & Aug-04-2009&04-Jul-2009(33.6)&05-Jul-2009(32.4) \\
    \hline
    \end{tabular}%
  \caption{Prediction of the critical time for both models. For each example, 225 time series
  are generated by varying the start time $t_1$ and end time $t_2$
  of the windows in which the calibration is performed. The mean
  value and the standard deviation of the predicted critical time $t_c$ among
  2250 predictions are shown in the table.}
  \label{tb:tcperform}%
\end{table}%

However, the new model makes it possible to determine the concentration of
stock gains over time from the knowledge of the Zipf factor. The two bubbles
are found to differ by the sign and contribution of the Zipf factor as well as
the factor load $\gamma$.

For bubble 1, the integrated Zipf factor $\zeta$ is positive as shown in Fig.
\ref{fg:fit_ssec_zipf1}, corresponding to the fact that valuation gains were
more concentrated on the large firms of the Shanghai index, especially in two
periods, Dec. 2006 - Jan. 2007 and Oct. 2007 - Dec. 2007. The factor load
$\gamma$ of the best fit in the example shown in Fig. \ref{fg:fit_ssec_zipf1}
is 0.44. And the statistics of $\gamma$ from all the 2250 fits of bubble 1 is
shown in the second row of Tab. \ref{tb:gamma}. All these results indicate
that the Zipf factor load $\gamma$ in bubble 1 is statistically large and
positive. This implies the existence of a lack-of-diversification premium that
contributes significantly to the overall price level in addition to the bubble
component.

A possible interpretation of the important of the Zipf factor is based on the
importance that investors started to attribute to the role of large companies
in driving the appreciation of the SSEC index during the first bubble. The
so-called 80-20 rule started to be hot among investors in discussions and
interpretation of the rising SSEC index. It was widely pointed out that the
growth of the SSEC index was driven essentially by 20\% of the stocks while the
other 80\% constituents of the index remains approximately flat (known as the
80-20 quotation of the Chinese stock market
\footnote{http://www.hudong.com/wiki/\%E4\%BA\%8C\%E5\%85\%AB\%E8\%A1\%8C\%E6\%83\%85}).
It is plausible that the widespread acknowledgement of the 80-20 rule led many
investors to discount the risk of a lack of diversification, therefore
enhancing the role of the Zipf factor.  This is consistent with our observation
that the Zipf factor load $\gamma$ is large and positive during the first
bubble period.

\begin{table}[h]
  \centering
    \begin{tabular}{|l|l|l|l|}
    \hline
    Example & $Mean of \gamma$& Median of $\gamma$&std of $\gamma$\\
    \hline
    Bubble 1 & 0.35&0.56&0.43\\
    Bubble 2 & -0.14&-0.11&0.15\\
    \hline
    \end{tabular}%
  \caption{Statistics of the Zipf factor load $\gamma$ from 2250 fit results.
  Most of the values for $\gamma$ for the period
  during the development of bubble 1 are positive and their average value is large.
  This means that the Zipf factor plays an important role during the
  development of bubble 1. The concentration
  of the stock market on a small number of large firms has a
  significant impact on the price change of
  the stock index.  In contrast, for bubble 2, the average value of $\gamma$ is
  relatively small and the exposition to the risk associated with a lack of diversification is found to
  be insignificant in pricing the value of the market.}
  \label{tb:gamma}%
\end{table}%

In contrast, the integrated Zipf factor $\zeta$ remained negative over the
lifetime of bubble 2 as shown in Fig. 2, implying that the gains of the
Shanghai index were more driven by small and medium size firms. The factor load
$\gamma$ is -0.028 for the best fit shown in Fig. \ref{fg:fit_ssec_zipf2} and
the mean value of $\gamma$ for bubble 2 is small and negative (see Tab.
\ref{tb:gamma}). The overall contribution of the Zipf factor to the stock
change is therefore small and negative (due to the product
of a negative integrated Zipf factor by a negative factor loading), which makes the remuneration of
investors due to their exposition to the diversification risk still positive
but small.

At the time when bubble 2 started, the world economy has been seriously shaken
by the developing subprime crisis. The demand for Chinese product exports decreased
dramatically. To compensate for the loss from collapsing exports, the Chinese government
launched a 4 trillion Chinese yuan stimulus with the aim to boost the domestic
demand. Small companies that are usually more vulnerable to a lack of access
to capital profited proportionally more than their larger counterpart from this
injection of capital in the economy. This is reflected in relative better performance
of small and medium size firms in the stock market, leading to
a slightly negative value of the integrated Zipf factor $\zeta$ during the
development of bubble 2. Although the small companies benefit more, the
stimulus was designed to boost the whole economy. The diversification risk
turned out to be relatively minor at that time, explaining the small value of the
Zipf factor load.

\section{Conclusion}
\label{sec:conclusion}

We have introduced a new model that combines the Zipf factor
embodying the risk due to lack of diversification with the Johansen-Ledoit-Sornette model
of rational expectation bubbles with positive feedbacks. The
new model keeps all the dynamical characteristics of a bubble described in the
JLS model. In addition, the new model can also provide information about
the concentration of stock gains over time from the knowledge of the Zipf
factor. This new information is very helpful to understand the risk
diversification and to explain the investors' behavior during the bubble
generation. We have applied this new model to two famous Chinese stock bubbles and
found that the new model provide sensible explanation for the diversification
risk observed during these two bubbles.

\clearpage

\appendix
\section{Derivation of the model with stochastic pricing kernel theory}
We present another derivation of the model using the theory of the
stochastic pricing kernel. Our derivation follows and adapt that
presented by Zhou and Sornette \cite{zhouSornette06}.

Under this theory, the no-arbitrage condition is presented as follows.
The product of the stochastic pricing kernel (stochastic
discount factor) $D(t)$ and the value process $p(t)$, of any
admissible self-financing trading strategy implemented by trading on
a financial asset, should be a martingale:
\begin{equation}
  D(t)p(t) = {\rm E_t}[D(t')p(t')], ~~~~~~\forall t'>t~.
\label{eq:pmmartingale}
\end{equation}

Let us assume that the dynamics of the stochastic pricing kernel is formulated as:
\begin{equation}
  \frac{dD(t)}{D(t)} = -r(t)dt - \gamma z(t)dt - \lambda(t)dW + \nu d \hat{W}~,
\label{eq:sdf}
\end{equation}
where $r(t)$ is the interest rate and $z(t)$ is the Zipf factor defined as
(\ref{hjyjuj4u}). The process $\lambda(t)$ denotes the market price of risk, as
measured by the covariance of asset returns with the stochastic discount factor
and $d \hat{W}$ represents all other stochastic factors acting on the
stochastic pricing kernel. By definition, $dW$ is independent to $d\hat{W}$ at
any time $t \geq 0$:
\begin{equation}
  {\rm E_t} [dW \cdot d\hat{W}] = {\rm E_t} [dW] \cdot {\rm E_t} [d\hat{W}] = 0~, \forall t \geq 0.
\end{equation}
We further use the standard form of the price dynamics in the JLS
model \cite{jls,jsl,js}:
\begin{equation}
  \frac{dp}{p} = \mu dt + \sigma(t)dW - \kappa dj~,
\label{eq:appendixprice}
\end{equation}
where $W$ is the same Brownian motion as in (\ref{eq:sdf}). The term $dj$ represents
the jump process, valued 0 when there is no crash and 1 when the crash
occurs. The dynamics of the jumps is governed by the crash hazard
rate $h(t)$ defined in (\ref{eq:hazard}) with:
\begin{equation}
  {\rm E}_t[dj] = h(t)dt~.
\end{equation}

According to the stochastic pricing kernel theory, $D \times p$ should be a
martingale. Taking the future time $t'$ in (\ref{eq:pmmartingale}) as
the increment of the current time $t$, then
\begin{eqnarray}
  {\rm E} \left[ \frac{p(t+dt)D(t+dt)-p(t)D(t)}{p(t)D(t)}\right] &=&
  {\rm E} \left[
  \frac{(p(t)+dp)(D(t)+dD)-p(t)D(t)}{p(t)D(t)}\right]\\\nonumber
  &=&{\rm E} \left[
  \frac{p(t)dD+D(t)dp + dD dp}{p(t)D(t)}\right]\\\nonumber
  &=& {\rm E} \left[ \frac{dD}{D} +
  \frac{dp}{p}+\frac{dDdp}{Dp}\right]\\\nonumber
  &=& 0~.
\end{eqnarray}
To satisfy this equation, the coefficient of $dt$ should be zero, that is
$ -r(t)+\mu(t) + \gamma z(t) +  \kappa h(t) + \sigma(t) \lambda(t) =0$.
This yields
\begin{equation}
\mu(t) =   r(t)+\gamma z(t) - \kappa h(t) - \sigma(t) \lambda(t) ~.
\end{equation}

When there is no crash ($dj = 0$), the expectation of the price
process is obtained by integrating (\ref{eq:appendixprice}):
\begin{equation}
  {\rm E_t} \left[ \ln p(t) \right] = \int (\gamma z(t)+\kappa
  h(t)+r(t)+\sigma(t)\lambda(t))dt~.
\end{equation}
For $r(t) = 0$ and $\lambda(t) = 0$, we obtain:
\begin{eqnarray}
  {\rm E_t} \left[ \ln p(t) \right] &=& \int (\gamma z(t)+\kappa
  h(t))dt \\\nonumber
  &=& \gamma \zeta(t) + \int \kappa h(t) dt \\\nonumber
  &=& \gamma \zeta(t) +  A + B(t_c-t)^m + C(t_c-t)^m\cos(\omega\ln (t_c-t) -
  \phi)~,
\end{eqnarray}
which recovers (\ref{eq:lppl}).

\clearpage

\clearpage

\end{document}